\documentclass[runningheads]{llncs}

\usepackage{graphicx}

\usepackage{xcolor}
\makeatletter
\newcommand{\printfnsymbol}[1]{
  \textsuperscript{\@fnsymbol{#1}}
}
\makeatother
\begin{document}

\title{Edge, Structure and Texture Refinement for Retrospective High Quality MRI Restoration using Deep Learning}

\author{Hao Li\inst{1}\thanks{These two authors contributed equally. Order is alphabetical} \and
Jianan Liu\inst{2}\printfnsymbol{1}}

\institute{University Hospital, Rupert-Karls-University Heidelberg, Heidelberg, Germany\\
\email{hao.li@med.uni-heidelberg.de}\\
\and
Derimis Tech., Gothenburg, Sweden\\
\email{chisyliu@hotmail.com}
}

\maketitle
\begin{abstract} Shortening acquisition time and reducing the motion-artifact are two of the most critical issues in MRI. As a promising solution, high-quality MRI image restoration provides a new approach to achieve higher resolution without costing additional acquisition time, modification on the pulse sequences or repeating the acquisition. Recently, as to the rise of deep learning, convolutional neural networks(CNNs) have been proposed to generate super-resolution images and reduce the motion-artifact for MRI applications. Recent studies suggest that using perceptual feature space loss and k space loss to capture the perceptual information and high-frequency information of images, respectively. However, the quality of reconstructed super-resolution and motion-artifact reduced MR image is limited cause the most important details of the informative area in MR image, the edges and the structure, can not be very well generated. Besides, lots of the super-resolution approaches are trained by using low-resolution images generated by applying bicubic or blur-downscale degradation, which can not represent the real process of MRI measurement. Such inconsistencies lead to performance degradation in the reconstruction of super-resolution MR images as well. This study reveals that using the L1 loss of SSIM and gradient map edge quality loss could force the deep learning model to focus on studying the features of edge and structure details of MR image, thus generating super-resolution MR image with more accurate, fruitful information and MR image with reduced motion-artifact. We employed a state-of-the-art model, RCAN, as the basic network framework into both MRI super-resolution and motion-artifact reduction tasks, trained the model by using low-resolution MR images and motion-artifact effected MR images which were generated by emulating how they are measured in the real world to ensure the model can be easily applied in the practical clinic environment, and verified the trained model could work fairly well.

\keywords {MRI \and Super Resolution Reconstruction \and Motion-Artifact Reduction \and Deep Learning \and Edge, Structure and Texture Refinement}
\end{abstract}
\section{Introduction}
\subsection{Background and Related Work}
Magnetic Resonance Imaging (MRI) nowadays is widely used in various medical applications to assist doctors to make accurate diagnoses, while doctors usually have to face the resolution-acquisition time trade-off problem in practice. The longer acquisition time leads to MR images with higher resolution. However, patients can not stay stationary for a long time without moving body. In contrast, The shorter acquisition time is more applicable for patients, whereas hardly generate MR image with sufficient resolution. Deep learning-based super-resolution reconstruction technology for MRI technology, which generates the high-resolution MRI images from low-resolution ones, could be used as a practical solution to provide high-quality MRI images without long acquisition time. Besides, patient movement during MRI measurement is inevitable, which generates motion-artifact and severely affects the image quality. Deep learning techniques can be used for motion-artifact reduction as well. It is also worth to mention such technologies could be applied easily in the clinical environment due that modification on the pulse sequences of the MRI machine is not needed.

Deep learning technology has been applied widely on variant tasks in computer vision since its first success on image classification tasks in 2013, drawing lots of attention from both academics and industries. Super-resolution image reconstruction, as one of the major fields of computer vision, was deeply influenced by deep learning technology. Deep learning-based super-resolution image reconstruction is a data-driven approach, collecting a big amount of image pairs which consist of low-resolution images and corresponding high-resolution images as training data, and training the deep neural network to extract the pixel-wise features and generate the up-sampled super-resolution image by using the pairs of training data.

Motion-artifact reduction is a comparable task to the super-resolution reconstruction for MR images, which restore the missing information of the images. Therefore, deep neural networks with similar structures and training strategies can be applied in both tasks.

The prior research in deep learning-based super-resolution visual image reconstruction, SRCNN\cite{ref_srcnn}, proposed to use bicubic as interpolation-based up-sampler to create the up-sampled low-resolution image first, then employed several convolutional layers to generate super-resolution image. With the invention of ResNet\cite{ref_resnet} and DenseNet\cite{ref_densenet}, residual block structure was applied by VDSR\cite{ref_vdsr} and dense block structure was incorporated by SRDenseNet\cite{ref_srdensenet}. Early study\cite{ref_perceptual_loss} introduced perceptual loss which used pre-trained VGG network to extract high-level features to represent the semantic and style information for super-resolution reconstruction. Some recent studies like EDSR\cite{ref_edsr}, DBPN\cite{ref_dbpn}, started to use sub-pixel convolutional layer\cite{ref_sub_pixel_conv}, which is a learnable up-sampling layer to generate the up-sampled super-resolution image in better quality by reshaping the generated channels by convolution.

Other studies have also revealed the possibility of implementing deep learning-based super-resolution algorithms on MR images, including neuroimaging applications\cite{ref_neuro_app}\cite{ref_neuro_app3}\cite{ref_neuro_app2}\cite{ref_neuro_app4}, cardiac imaging\cite{ref_cardio_app} and musculoskeletal imaging\cite{ref_mus_app}, and obtained promising results. Besides, motion-artifact reduction has also been performed in some studies using widely-used deep neural networks for computer vision tasks\cite{ref_mar1}\cite{ref_mar2}. 

It is also interesting to note that, prior studies applied different types of loss functions to train the neural networks towards specific feature refinements. Pixel-wise loss is the most commonly used loss function, which is employed in all studies to minimize the pixel-to-pixel difference between the reconstructed and the target images\cite{ref_neuro_app}\cite{ref_cardio_app}. Frequency domain loss was introduced in the study of DAGAN for artifact reduction of MR image\cite{ref_kspace}. More importantly, structure similarity index (SSIM)\cite{ref_ssim}, as one of the most popular image quality assessment index, is used in some most recent studies to drive the network restored images towards better similarity in structures with ground truth (GT) images\cite{ref_cardio_app}.

\subsection{The Contributions of Our Proposed Method}
In our study, the following contributions are made:
\begin{itemize}
    \item 1). We proposed a low-resolution MRI image generation approach with cropping and disposing of high-frequency components in the frequency domain instead of conventional downsampling methods in computer vision tasks.
    \item 2). We utilized residual channel attention network (RCAN)\cite{ref_rcan} which has been proved to provide outstanding performance in super-resolution tasks for the visual image, as the backbone of our neural network model for generic high-quality MRI reconstruction task: MRI super-resolution and MRI motion-artifact reduction. 
    \item 3). We also explored the progressive structure of the network based on RCAN for super-resolution with high SR factors.
    \item 4). Besides, we implemented pixel-wise Charbonnier loss, MSE loss of frequency domain, and L1 loss of SSIM map and gradient map for the refinement of anatomical structures.
    \item 5). Furthermore, we also proposed further processing to focus on reconstruction of the high-frequency components in the frequency domain and small gradients in gradient map, which would benefit the recovery of edge and texture for MRI image.
\end{itemize}

The experiments were taken using pixel-wise Charbonnier loss only or together with other types of loss functions, and the results were compared based on SSIM and PSNR to show the benefits from the proposed loss functions.

\section{Method and Experiments}
In the MRI super-resolution reconstruction task, we first generated the lower resolution (LR) MRI image from measured high resolution (HR) MRI image by mimicking how the LR images are measured on the real machine, cropping each LR-HR pair into several 2D image patches in the same size, divided the image patches in training, validation, and test datasets, and augment the patches in the training dataset. Secondly, we trained the network models using the training dataset, to minimize the selected loss functions with different weights via back-propagation of gradient. Then we examined the performance of the trained network model by using other batches of LR-HR patches that have never been used in the training phase, to identify the best network model to reconstruct the MRI image. Subsequently, we generate the reconstructed MRI image by feeding the LR patches into a well-trained network model from the previous step. 

In the MRI motion-artifact reduction task, we also followed the strategy of mimicking the generation of real motion-artifact. The motion-artifact is caused by the movement of the patient during the measurement, which leads to the incorrect phase encoding in certain k-space lines. We employed the method that generates the affected k-space lines by applying FFT on shifted or rotated images and using segments from k-spaces of these moved images to replace the corresponding k-space segments of the original images. The data pairs of MRI images with and without motion-artifact are fed into our network model for training and the trained network model is used to generate the reconstructed MRI image whose motion-artifact is removed.

The details of our method are described in the following sections.

\subsection{High Quality MRI Image Restoration}
The high-quality image restoration task refers to design a restoration system function, \textit{h}, which inputs low quality (LQ) image(e.g. low-resolution image, or image with blur, motion-artifact, etc.) and outputs the corresponding restored image(e.g. SR image or image without blur, motion-artifact) in high quality (HQ) as shown in the following equation.
\begin{equation}
HQ = h(LQ)
\end{equation}
For super-resolution task, the LQ image is generated from HR image through resolution degradation system function, \textit{f}, which usually represents a particular down-sampling or performance degradation function, \textit{f}, as below:
\begin{equation}
LQ = f(HR)
\end{equation}
As can be seen, in order to make the reconstructed SR image as close to HR image as possible, the task of designing \textit{h} is actually equivalent to find the inverse of function \textit{f}:
\begin{equation}
HQ = h(LQ) = f^{-1}(LQ)
\end{equation}
Prior studies showed that such task is an ill-pose inverse problem, the analytical solution of the inverse of \textit{f} does not exist. The researches in computer vision revealed CNN based models have the great potentials to fit the inverse of \textit{f} in the generic computer vision domain. Consequently, the CNN based models have been investigated to solve both MRI super-resolution image reconstruction and MRI motion-artifact reduction in recent studies. 

However, MRI super-resolution image reconstruction still has fundamental differences compared to generic computer vision application, which makes it difficult to apply the existing CNN based super-resolution reconstruction network model into the MRI field directly. One of the major difference is, the common approach to down-sampling the HR visual image to generate the LR visual image in computer vision application could be expressed as:
\begin{equation}
LR = f(HR) = \theta(HR) + \epsilon
\end{equation}
where $\theta$ denotes a Gaussian blurring function, and $\epsilon$ represents the Gaussian noise, most of the studies in MRI super-resolution reconstruction follow the same approach to generate the LR MRI image. On the contrary, the measured LR MRI image in the real measurements does not follow the similar procedure at all. The super-resolution reconstruction model trained and evaluated by using the LR MRI images generated from the same approach, will be hard to use in the real world for doctors to give accurate diagnostics. To reduce this gap, we abandoned such down-sampling approach commonly used in computer vision application, designed a new approach which can be seen as an emulation of real LR MRI image measurements. For MRI image motion-artifact reduction, we also applied an approach to add the motion-artifact in the way of simulating the procedure of real MRI image measurements with motion-artifact\cite{ref_mar2}. 

Moreover, there is much less semantic information that could be extracted in MRI image, especially for the low-frequency components which represent the texture information. In addition, doctors concern more on the accuracy of the reconstructed high-quality MRI image, particularly on the edge area of different tissues rather than the perceptive visual feeling. We refine the network model by enhancing the information perception and extraction capabilities in the texture and edge area for high-quality MRI image reconstruction. We will explain the details of how these issues have been solved by our method in the later sections.

\subsection{RCAN based High Quality MRI Image Reconstruction}
Residual Channel Attention Network(RCAN), has been proposed recently\cite{ref_rcan} to generate the SR visual image from LR visual image in the generic computer vision field. Similarly, in the major part of the RCAN framework belongs to our model, a dedicated module called channel attention(CA) layer is embedded into every residual block\cite{ref_resnet} to form a new module, residual channel attention block(RCAB). Several RCABs construct one residual group and a long skip connection, and the same pattern has also been extended to form residual in residual(RIR) module, consists of several residual groups(RGs) and a long skip connection as well. The attention weights for different channels that carry the statistics of semantic information of MR image feature map will be learned and used to force the network model to focus on restoration of the information which has more important semantic information, meanwhile, the skip connection in RCAB, RG, and RIR could provide the feasibility of training the network model in an easy pace, especially makes the low-frequency component information bypassed directly that enforces the RCAN focus on learning high-frequency component information. All the weights of RCAN based SR MRI network model are trained by back-propagation, driven by minimization of the primary and refinement loss functions represent differences between the predicted HQ MRI image and ground truth HR MRI image. The up-sampling module performs the up-scaling operation on learned feature maps, results in the SR image which has the same size as the HR image when implements the MRI super-resolution task.

In our experiment of MRI SR task, the HR images had a larger size compared with the generated LR images along with both height and width. We constructed the RCAN model with $N$ RGs, $M$ RCABs per RG, and $K$ 2x up-sampling module to utilize up-scaling of MRI image, to load the cropped 2D LR images with only one channel, generate corresponding SR images, and piece several SR MRI images together into a complete reconstructed SR image. For the MRI motion-artifact reduction task, the MRI images with motion-artifact (MA) have the same size as the MA reduced (MAR) images, thus we choose to insert a down-sampling module before the up-sampling module for the MRI motion-artifact reduction task, to keep the same size of MA images and MAR image while applying larger perceptive field for learning multiple-scale feature. The basic pipeline of our model for high-quality MRI reconstruction is shown in Figure 1.

\begin{figure}
\includegraphics[width=\textwidth]{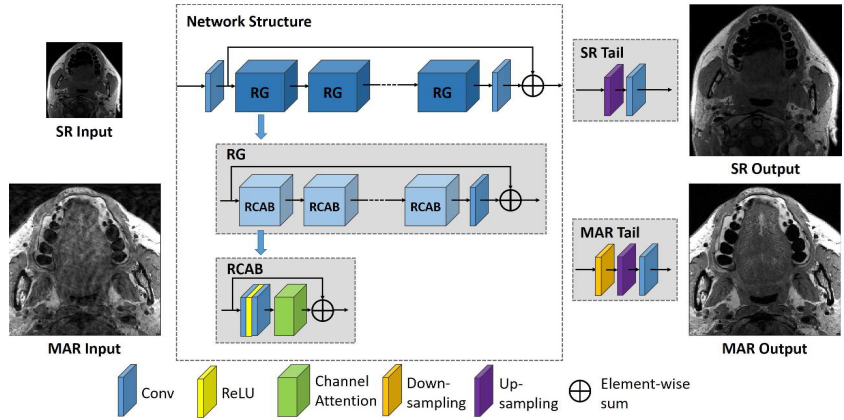}
\caption{The entire pipeline of RCAN based MRI SR and MAR network model} \label{Fig_1}
\end{figure}

\subsection{Progressive-upsampling network for high up-scaling SR factor}
Considering the original up-scaling approach in the RCAN network, for large up-scaling factor, e.g. 4x, 8x, is simply appending $K$ pixel shuffle module with up-scaling factor 2 consecutively at the end of the network model to utilize $2^K$ times up-sampling. We argue that such a simple approach loses the capability to exact and get used to the semantic features with different perceptive fields, it will have a hard time to get hold of structure consistency and semantic information accuracy for the reconstructed SR image when the up-scaling factor is large. Thus we also propose to use progressive-upsampling RCAN network to replace the original RCAN framework for high up-scaling factor, e.g. 4x, 8x. In the progressive-upsampling network framework, the $K$ stage of the original RCAN are connected consecutively, one pixel shuffle module with up-scaling factor 2 at the end of each stage. The details of two different network frames could be seen in Figure 2.

\begin{figure}
\includegraphics[width=\textwidth]{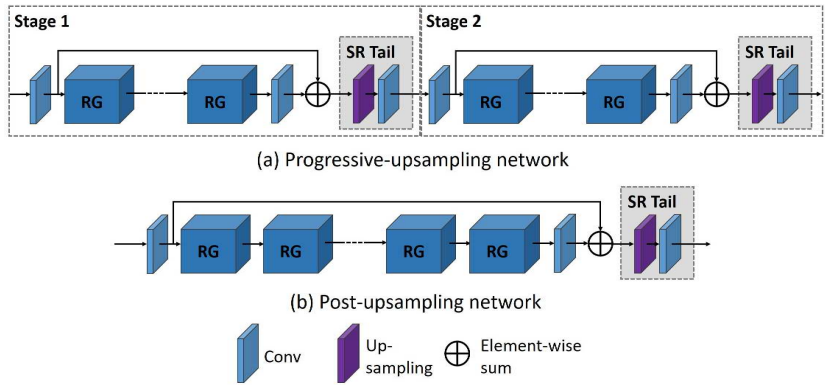}
\caption{Comparison of structures of progressive and post-upsampling networks with high SR factor.} \label{Fig_2}
\end{figure}

\subsection{Frequency, Edge, Structure and Texture Refinement}
Prior studies applied different types of loss functions to train the neural networks towards specific feature refinements. Pixel-wise L1 loss is the most commonly used loss function, which is employed in nearly all studies to minimize the pixel-to-pixel difference between the reconstructed and the target images. Pixel-wise Charbonnier loss was used in this study as a differentiable l1 loss to avoid a strong smoothing effect\cite{ref_charbo}. Frequency domain loss was introduced in the study of DAGAN for artifact reduction of MR image\cite{ref_kspace}. In this study, we also involved frequency domain MSE loss and applied additional processing for the refinement of high-frequency components. Furthermore, SSIM loss is used in some most recent studies to drive the network reconstructing the high-quality image towards better similarity in structures with HR/GT image. It is normally applied as 1 subtracted by SSIM value measured between MRI SR/MAR images and HR/GT images \cite{ref_cardio_app}. In this study, we utilized the L1 loss of the SSIM index. At last, the gradient map loss is not frequently used in other studies but helps to restrain the boundaries between tissues. Therefore, it is utilized in this study as the L1 loss calculated between the gradient map of SR image/ MAR image and HR/GT image. Besides, we applied further amplification on small values in the gradient maps for the refinement of textures.

To enable the capability of restoring the SR/MAR image, the pixel-wise L1 Charbonnier loss which counts summation of absolute errors over every pixel between SR/MAR image and HR/GT image is used in our model as a key component in the primary loss function. The Charbonnier loss is considered to guide the network model to reconstruct high-quality MRI images without too much over-smoothing phenomenon. 

Despite the primary loss function could provide fairly good guidance in generic SR image reconstruction task for computer vision application, the MRI SR/MAR image reconstruction task still suffer lots of problems as we pointed out in the previous section. Accordingly, we propose several other loss functions and support approaches to refine the reconstruction capability in edge, structure, and texture. Particularly, we obtain the k-space representation, gradient map, and SSIM for both reconstructed MRI high-quality image and ground truth image, then calculate mean squared error(MSE) loss for k-space representation, the L1 loss for gradient map and SSIM between SR/MAR images and HR/GT images respectively, as our refinement loss to refine the information represented in frequency, edge, and structure.

Besides, to overcome the shortage of semantic information in texture, we also propose to amplify the small values in the calculated gradient map by the following equation:
\begin{equation}
M = 1 - e^{(-aM)}
\end{equation}
where M is the gradient map, \textit{a} is a hyper-parameter which we set up as 2.5, enforce such small values in gradient map to become relatively large, since the small value in gradient map represents the texture information. Thus the texture information could be refined. Such refinement can be seen clearly in the example shown in figure 3.

\begin{figure}
\includegraphics[width=\textwidth]{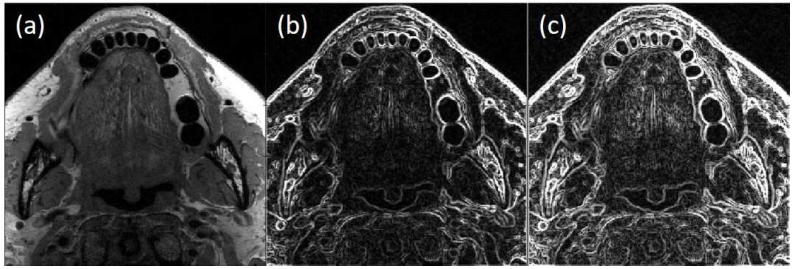}
\caption{Comparison of original and amplified gradient-map. (a): original image, (b): original gradient-map, (c): gradient-map with amplified small gradients.} \label{Fig_3}
\end{figure}

Moreover, we also amplified the high-frequency components in the k-space representation by multiplying with a 2D Gaussian distribution mask function. We expected the network model to focus on the high-frequency components but the values of high-frequency components are comparably lower than the low-frequency components. Thus the multiplication with Gaussian mask should minimize the imbalance between the different intensities of high and low-frequency regions in the k-space representation. To define a proper value for standard deviation, $\sigma$, of the Gaussian mask, we surveyed the Gaussian distribution with $\sigma$ values between 10 and 50 with an increment of 1, figured out the difference between averages and standard deviations of the intensities in low-frequency and high-frequency regions after the k-space multiplied with the Gaussian mask is minimized with $\sigma$ = 32 for the dataset used in our study. Beware in the k-space of MRI image, the high-frequency components are usually represented at four corners and low-frequency components exist in the central region of k-space representation. However, to apply the 2D Gaussian mask, we overturn the distribution of high and low-frequency components in our experimental data thus the high-frequency components are located in the center.

As shown on the lower row in Figure 4, the high-frequency components distributed in the central part of the k space representation are enhanced by such a 2D Gaussian mask, compared to the original k-space representation of the original MRI image shown on the upper row in the same figure.
\begin{figure}
\includegraphics[width=\textwidth]{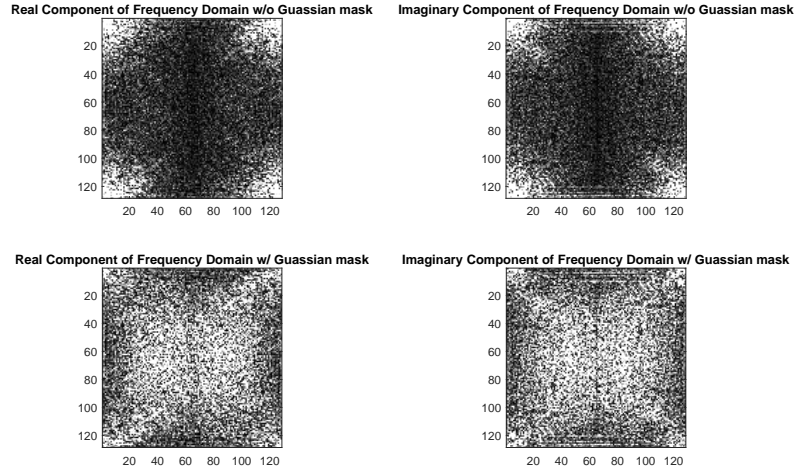}
\caption{The k space representation of MRI data without and with 2D Gaussian mask. Note the high frequency components are located in the central part in this figure} \label{Fig_4}
\end{figure}

Due to the high-frequency components in the frequency domain actually stand for the edge information in the image domain, the edge information will be refined further by using such a 2D Gaussian mask.

\subsection{Data acquisition and Pre-processing}

The measurements were performed on a 3-Tesla MRI system (MAGNETOM Trio, Siemens Healthcare, Germany) with a mandibular coil (NORAS MRI Products GmbH, Germany). Works-in-progress software packages by Siemens Healthcare provided the prototype sequence of MSVAT-SPACE. The sequence was optimized for clinical application with regards to reasonable image quality, acquisition time, and field-of-view. 

3D image datasets were acquired using MSVAT-SPACE, with PD weighting, 0.44mm isotropic resolution, and in-plane matrix size was 384x300. Each 3D dataset was cropped in the x-y plane to a 256x256 region-of-interest before down-sampling.  

Different from super-resolution tasks for visual images, where the low-resolu-tion images are generated using bicubic or blurring down-sampling, a Fourier down-sampling was performed to simulate the acquisition of low-resolution images with the smaller matrix. The images were transformed to the frequency domain with FFT, and the central region of k-space with varying sizes based on the down-sampling factor was retained and transformed back to the image domain with iFFT. For both high resolution and generated low-resolution images, the intensities of voxels were scaled to 0 and 1. An example of the generation of LR MRI images with scale factor 2 is shown in Figure 5.

\begin{figure}
\includegraphics[width=\textwidth]{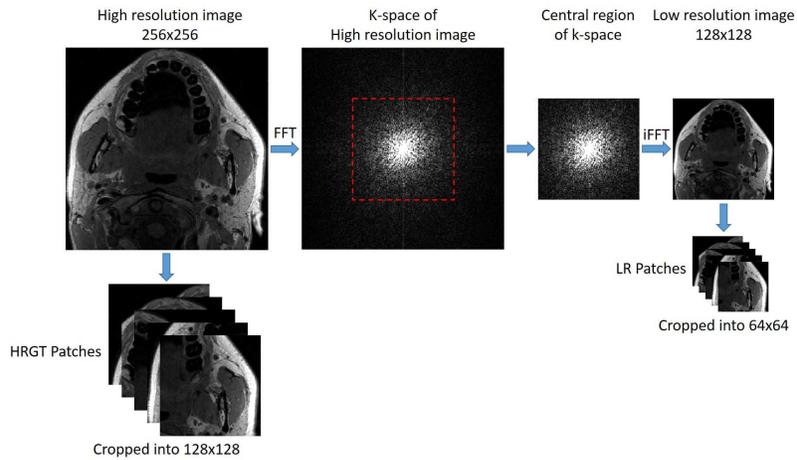}
\caption{Generation of LR MRI images with scale factor of 2 and patch cropping} \label{Fig_5}
\end{figure}

After down-sampling, the high-resolution images were cropped into 128x128 patches with a stride of 64, the low-resolution images were cropped into 64x64 patches with a stride of 32 for an SR factor of 2 and 32x32 patches with a stride of 16 for an SR factor of 4. Data augmentation was performed by rotating the patches by 90, 180, and 270 degrees. 

For a retrospective generation of motion-artifact in MR images, the method of splicing lines from multiple k-spaces was applied to mimic the generation of real motion-artifact. As shown in Figure 6, a group of images was generated from each slice of the original image by a linear shift in the range of 1 to 8 voxels and rotation in the range of 0 to 5 degrees. Then the original image and manipulated images were transformed to k-space using FFT, and k-space segments of the original image, which were randomly selected, were replaced with the segments from k-spaces of the manipulated images. The images for the motion-artifact correction task were not cropped, therefore the sizes of input MA images and GT images are both 256x256. Data augmentation was applied by generating 5 motion-artifact affected images from each original image. These 5 MA images had different patterns of movement and k-space segment replacement.

\begin{figure}
\includegraphics[width=\textwidth]{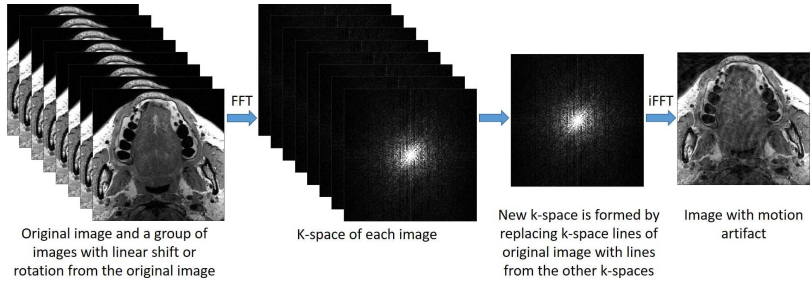}
\caption{Generation of MRI images with motion-artifact} \label{Fig_6}
\end{figure}

\subsection{Implementation Details}
This study involved 28 datasets, which were randomly divided into training group with 21 datasets (77472 patches for super-resolution task, 10760 images for motion-artifact correction), a verification group with 4 datasets (3600 patches for super-resolution task, 2000 images for motion-artifact correction), and test group with 3 datasets (2520 patches for super-resolution task, 1400 images for motion-artifact correction). Training datasets were used for training the neural network, while verification datasets were used to monitor the performance of the neural network during training, but were not involved in training. Test datasets were used for the final test of the trained network and were isolated from the training process.

As to the implementation of RCAN, for SR factor of 2, RG number of 5 was set, with 5 RCAB in each RG. Conv layers in shallow feature extraction and RIR structure have 64 filters, except for that in the channel-downscaling. For SR factor of 4, we connected two stages of the RCAN network sequentially for progressive-upsampling, consists of 5 RGs with 5 RCABs and one up-sampling module which up-samples the LR images with an SR factor of 2 at the end of each stage. And for post-upsampling, we adopted a single-stage RCAN network with 10 RGs and 5 RCABs in each RG, and up-sample the LR images with an SR factor of 4 directly at the end of the single-stage network.

The networks were trained on two workstations equipped with either two Quadro GV100 graphic cards or one Quadro P2000 graphics card (Nvidia, Santa Clara, Calif ). We used Pytorch 1.7 at the back end for all deep learning experiments. In each training batch, eight LR patches were randomly extracted as inputs. We trained our model 60 epochs using ADAM optimizer with $\beta_1 = 0.9$, $\beta_2 = 0.999$, and $\epsilon= 1e-8$, and learning rate was set with 5 epochs of linear warm-up, and afterwards as 1e-4 and divided by 2 every 10 epochs.

\section{Results}
\subsection{Super-Resolution with Different Loss Functions}
We evaluated the performance of super-resolution reconstruction for MR images using RCAN as the backbone with an SR factor of 2 and multiple combinations of loss functions. The combinations of loss functions are listed in Table 1. We compared the SSIM and PSNR of reconstructed images from 2x downsampled low-resolution images with L1 loss, Charbonnier loss, the combination 1 of four loss components (Charbonnier loss, SSIM L1 loss, k-space MSE loss, and gradient-map L1 loss), and the combination 2 of enhanced four loss components (Charbonnier loss, SSIM L1 loss, k-space+ MSE loss, and gradient-map+ L1 loss). Table 1 shows the quantitative comparison results. The results reveal that with the same network structure, different loss functions could help to refine the reconstructed super-resolution images in different aspects. Besides, the best result was achieved by combination 2, which means the amplification applied on k-space and gradient-map has enhanced weights of the low-frequency components in K-space loss and the gradient of textures in the gradient-map loss. 

\begin{table}
\caption{Effects of different Loss components. The best SSIM and PSNR (dB) values were achieved in 40 epochs. }\label{tab1}
\begin{tabular}{|l|c|c|c|c|}
\hline
Loss Components & R1 & R2 & R3 & R4 \\
\hline
Pixel-wise L1& Y &   &   &\\
Pixel-wise Charbonnier&   & Y & Y & Y\\
SSIM L1&   &   & Y & Y\\
K-Space MSE&   &   & Y &  \\
Gradient Map L1&   &   & Y &  \\
K-Space+ MSE&   &   &   & Y\\
Gradient Map+ L1&   &   &   & Y\\
\hline
SSIM & $0.9211\pm0.0175$ & $0.9212\pm0.0175$  & $0.9229\pm0.0176$ & $0.9236\pm0.0172$\\
PSNR & $34.29\pm1.82$ & $34.31\pm1.89$ & $34.38\pm1.82$ &  $34.54\pm1.88$\\
\hline
\end{tabular}
\end{table}
\begin{figure}
\includegraphics[width=\textwidth]{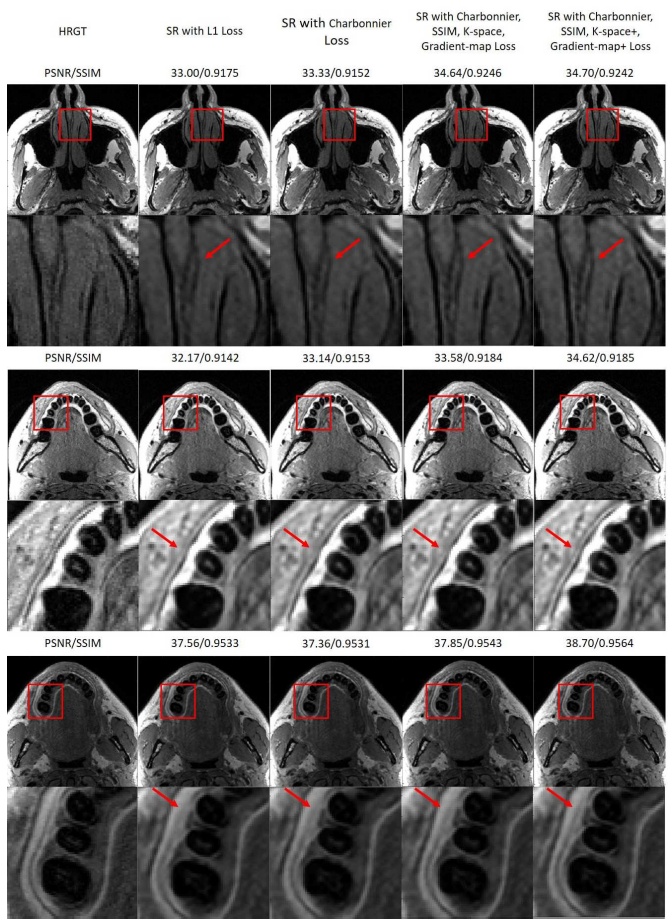}
\caption{Performance comparison of different loss functions with SR factor of 2.} \label{Fig_7}
\end{figure}

Figure 7 shows the results of super-resolution images reconstructed with different loss functions. it can be found that, besides improving SSIM and PSNR, SSIM loss, k-space loss, and gradient-map loss, also enhance the reconstruction of small structures with optimal sharpness and contrast at boundaries. And the amplification algorithms on the k-space and gradient map further improves the performance of depicting small structures and texture.

\begin{figure}
\includegraphics[width=\textwidth]{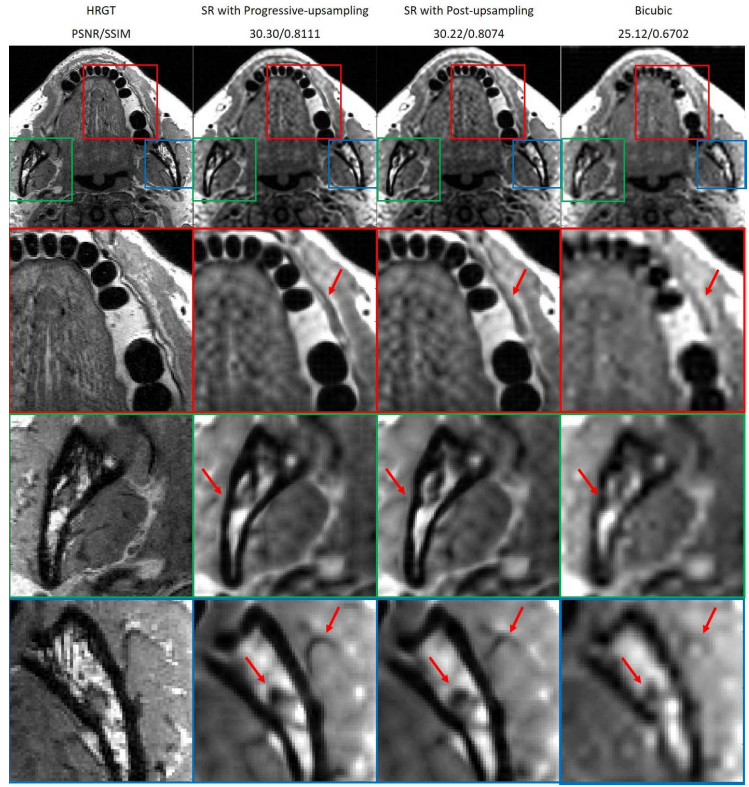}
\caption{Comparison between progressive- and post-upsampling with SR factor of 4.} \label{Fig_8}
\end{figure}

\subsection{Super-Resolution with Conventional Post-upsampling and Progressive-upsampling for High SR Factors}

Besides 2x reconstruction, we also performed super-resolution reconstruction with a higher SR factor with two different up-sampling schemes, the traditional post-upsampling scheme originally used in the RCAN network model and the progressive-upsampling scheme.

The two network schemes give comparable SSIM and PSNR values on the test dataset, which on average are SSIM of $0.8181\pm0.0288$ and PSNR of $29.90\pm1.60$ for progressive-upsampling, and SSIM of $0.8171\pm0.0294$ and PSNR of $29.91\pm1.59$ for post-upsampling. Although the difference in SSIM and PSNR is negligible, the progressive-upsampling network is highly superior to the post-upsampling network in the accuracy of the anatomical structure. As shown in Figure 8, arrows indicate significant differences in the reconstructed images, which reveal that the shapes of small anatomical structures in the images from the progressive-upsampling network are much more close to the ground truth than the post-upsampling network. However, we also observed a disadvantage of the progressive-upsampling network, which is the chessboard artifacts in some regions in the reconstructed super-resolution images. The chessboard artifacts were not observed in the super-resolution images reconstructed with the post-upsampling network.

\begin{figure}
\includegraphics[width=\textwidth]{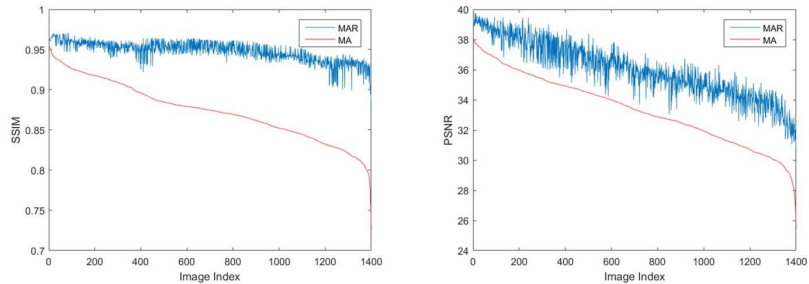}
\caption{SSIM and PSNR of MAR and MA images with different severity of motion-artifact.} \label{Fig_9}
\end{figure}

\begin{figure}
\includegraphics[width=\textwidth]{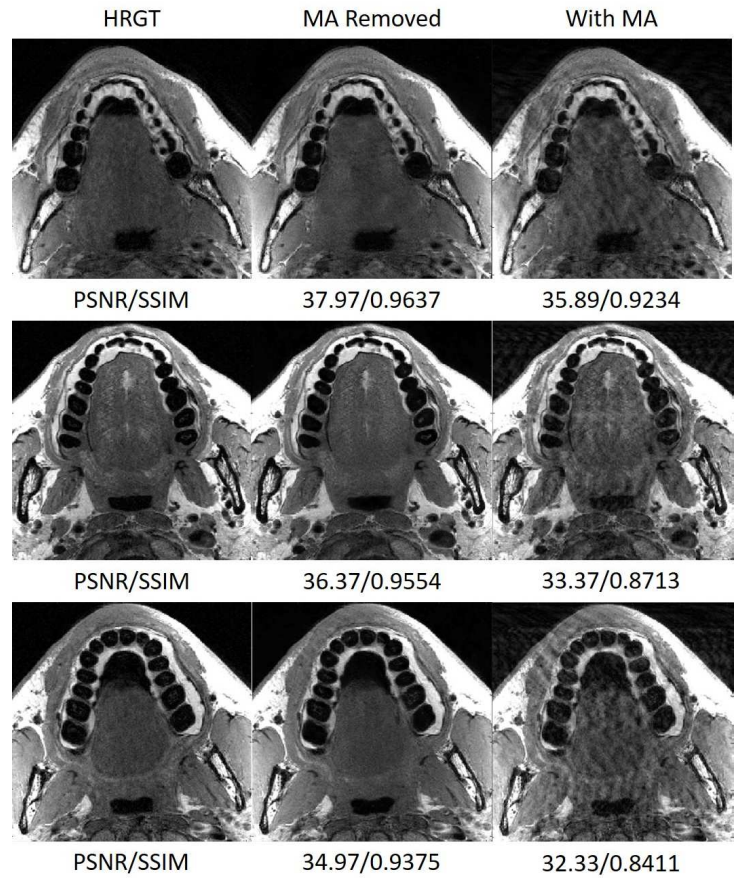}
\caption{Results of motion-artifact reduction.} \label{Fig_10}
\end{figure}

\subsection{Motion-Artifact Reduction}

Besides of reconstruction of super-resolution MR image, the proposed network is also applied to the task of motion-artifact reduction. Our proposed network model achieves excellent results, with significant improvements in SSIM and PSNR. For the MA images, the SSIM and PSNR are $0.8743\pm0.0361$ and $33.37\pm2.28$, respectively. After the process of motion-artifact reduction with the proposed network, the SSIM and PSNR rise to $0.9481\pm0.0124$ and $36.00\pm1.87$, respectively. Figure 9 shows the change of SSIM and PSNR for each pair of test images with different severity of motion-artifact. It can be found that the SSIM and the PSNR of MAR images are consistently higher than those of MA images. The MA images and MAR images have a comparable tendency in PSNR, and the SSIM of MAR images is very stable.

The examples in Figure 10 also show that, in visual effects, the majority of features have been kept and very well represented while the motion artifact is significantly reduced.

\section{Conclusion}
In this paper, we proposed to RCAN based generic network model to restore the high-quality MRI image. Particularly, we employed multiple loss functions with high-frequency components and small gradients refinement for recovering structure, edge, and texture information for both MRI SR task and MRI MA reduction task, achieved remarkable performance in both SSIM, PSNR, and visual quality. We also claimed the progressive scheme with our proposed model leads to a more accurate MRI SR image than the original post-upsampling scheme used in RCAN for higher SR factor, e.g. 4x, 8x, thus made MRI SR reconstruction technology more pragmatic to be used in the practical clinic environment.

\end{document}